\documentclass[reprint,superscriptaddress,showpacs,prl]{revtex4-2}
\pdfoutput=1
\usepackage{float}
\usepackage{color,graphicx}
\usepackage{hyperref}
\usepackage{mathrsfs}
\usepackage{bm}
\usepackage{amsmath,amssymb}
\usepackage{natbib}
\begin{document}
\title{Nearly room temperature ferromagnetism in pressure-induced correlated metallic state of van der Waals insulator CrGeTe$_3$}
\author{Dilip Bhoi}
\email{dilipkbhoi@issp.u-tokyo.ac.jp}
\affiliation{The Institute for Solid State Physics, University of Tokyo, Kashiwa, Chiba 277-8581, Japan}
\author{Jun Gouchi}
\affiliation{The Institute for Solid State Physics, University of Tokyo, Kashiwa, Chiba 277-8581, Japan}
\author{Naoka Hiraoka}
\affiliation{Department of Physics, Graduate School of Sciences, University of Tokyo, Tokyo 113-0033, Japan}
\author{Yufeng Zhang}
\affiliation{The Institute for Solid State Physics, University of Tokyo, Kashiwa, Chiba 277-8581, Japan}
\affiliation{School of Physics, Southeast University, Nanjing 211189, China}
\author{Norio Ogita}
\author{Takumi Hasegawa}
\affiliation{Graduate School of Integrated Arts and Sciences, Hiroshima University, Higashi-Hiroshima, Hiroshima 739-8521, Japan}
\author{Kentaro Kitagawa}
\affiliation{Department of Physics, Graduate School of Sciences, University of Tokyo, Tokyo 113-0033, Japan}
\author{Hidenori Takagi}
\affiliation{Department of Physics, Graduate School of Sciences, University of Tokyo, Tokyo 113-0033, Japan}
\affiliation{Institute for Functional Matter and Quantum Technologies, University of Stuttgart, 70569 Stuttgart, Germany}
\affiliation{Max Planck Institute for Solid State Research, Heisenbergstra\ss e 1, 70569 Stuttgart, Germany}
\author{Kee Hoon Kim}
\affiliation{Department of Physics and Astronomy, CeNSCMR, Seoul National University, Seoul, 151-747, Republic of Korea}
\affiliation{Institute of Applied Physics, Seoul National University, Seoul, 151-747, Republic of Korea}
\author{Yoshiya Uwatoko}
\affiliation{The Institute for Solid State Physics, University of Tokyo, Kashiwa, Chiba 277-8581, Japan}
\date{\today}
\begin{abstract}
A complex interplay of different energy scales involving Coulomb repulsion, spin-orbit coupling and Hund's coupling energy in two-dimensional (2D) van der Waals (vdW) material produces novel emerging physical state. For instance, ferromagnetism in vdW charge transfer insulator CrGeTe$_3$, that provides a promising platform to simultaneously manipulate the magnetic and electrical properties for potential device implementation using few layers thick materials. Here, we show a continuous tuning of magnetic and electrical properties of CrGeTe$_3$ single crystal using pressure. With application of pressure, CrGeTe$_3$ transforms from a FM insulator with Curie temperature, $T_{\rm{C}} \sim $ 66 K at ambient condition to a correlated 2D Fermi metal with $T_{\rm{C}}$ exceeding $\sim$ 250 K. Notably, absence of an accompanying structural distortion across the insulator-metal transition (IMT) suggests that the pressure induced modification of electronic ground states are driven by electronic correlation furnishing a rare example of bandwidth-controlled IMT in a vdW material.
\end{abstract}
\maketitle
\newpage
Discovery of two-dimensional (2D) magnetism in van der Waals (vdW) materials have unfurled diverse range of possibilities for development of novel spintronics, multiferroic and quantum computing devices using atomically thin material as well as fundamental research \cite{Huang2020,Burch2018,Gong2019,Zhong2017,Yao2019,Seyler2018}. These also include exploration of exotic physics such as Kitaev quantum spin liquid state in 3$d$ electron system with $S$ = $3/2$ \cite{Xu2019,Xu2020}. Until now, Cr$Y$Te$_3$ ($Y$ = Ge, Si) \cite{Carteaux1995,Gong2017}, Cr$X_3$ ($X$ = I and Br) \cite{McGuire2015,Tsubokawa,Huang2017,Kim2019c,Zhang2019a}, 1$T$-CrTe$_2$ \cite{Freitas2005,Zhang2021} and Fe$_{3-x}$GeTe$_2$ \cite{Deiseroth2006,Chen2013,Deng2018,Fei2018}, are the few vdW materials known to exhibit ferromagnetic (FM) order in bulk single crystal form and retain intrinsic ferromagnetism down to monolayer limit. Among these, Cr$Y$Te$_3$ and Cr$X_3$ are Mott insulators with a charge gap, facilitating a suitable platform to exploit both charge and spin degrees of freedom. 

\begin{figure}[th]
\includegraphics[width=0.47\textwidth]{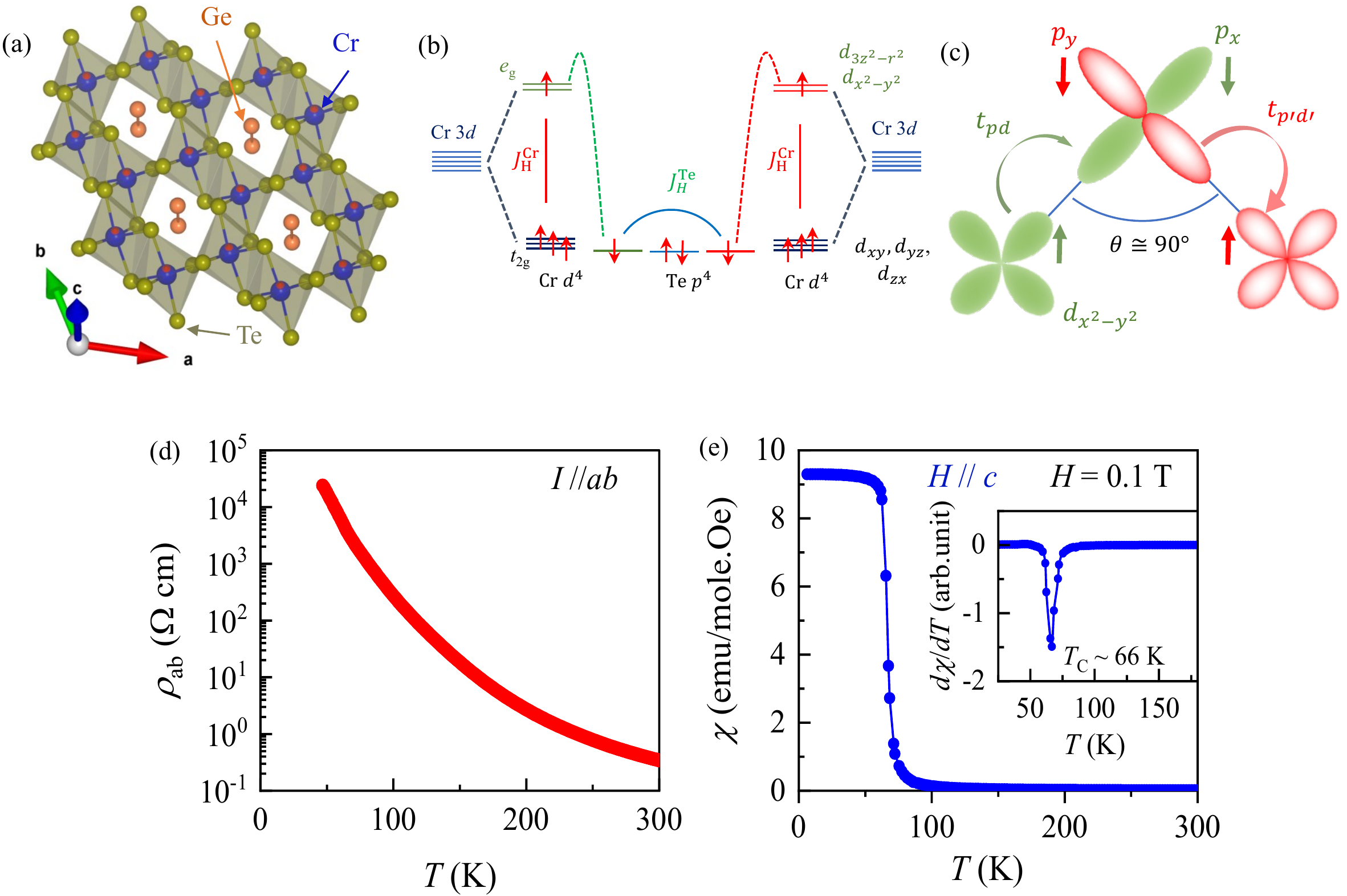}
\caption{(a) Single layer of CrGeTe$_3$ illustrating the honeycomb network of edge sharing CrTe$_6$ octahedra. In case of Cr$X_3$, the place of Ge-Ge dimers remains vacant. (b) Crystalline electric field splitting of Cr 3$d$ orbitals into $t_{2g}$- and $e_g$-manifolds. (c) Schematic picture of indirect superexchange FM interaction between Cr $e_{g}$-orbitals via two Te $p$-orbitals, where $J_{\rm{H}}^{\rm{Cr}}$ and $J_{\rm{H}}^{\rm{Te}}$ are respectively the Hund\textquoteright s coupling at Cr and Te site. $t_{pd}(t_{p'd'})$ is the virtual hopping between $e_{g}$($e_g'$) and $p$($p'$) orbitals. (d)Temperature dependence of the in-plane resistivity, $\rho_{ab}$, with current flowing in the $ab$-plane. (e) Zero field-cooled (ZFC) dc magnetic susceptibility, $\chi$, measured at field $H$ = 0.1 T applied along the easy magnetization $c$-axis. Inset shows the temperature derivative, $d\chi/dT$.}\label{SE_M}
\end{figure}

At low temperature, these Cr-based vdW insulators share a common layered rhombohedral $R\bar{3}$ crystal structure held together by weak vdW forces along the $c$-axis \cite{McGuire2015,Tsubokawa,Carteaux1995}. Each single layer consists of a honeycomb network of edge sharing octahedra formed by a central Cr atom bonded to six ligand atoms (Te or $X$), as illustrated in Fig.\ref{SE_M}(a) for CrGeTe$_3$ as a representative. Crystalline field effect ensuing from this octahedra splits Cr-3$d$ orbitals into $t_{2g}$- and $e_g$-manifolds [Fig.\ref{SE_M}(b)]. The onsite Coulomb repulsion localizes the $t_{2g}$-electrons driving the system into an insulating state significantly well above the Curie temperature, $T_{\rm{C}}$ \cite{Kang2019,Zhang2019,Suzuki2019b}. Although, a direct antiferromagnetic (AFM) exchange interaction exists between $t_{2g}$ electrons, thermal fluctuation inherent to 2D suppress the long-range magnetic order. The spin orbit coupling (SOC) emanating through the covalent bond between ligand $p$ and Cr-$e_g$ orbital generates the magnetocrystalline anisotropy energy (MAE) to counteract the thermal fluctuation \cite{Kim2019,Zhang2019}. Below $T_{\rm{C}}$, superexchange interaction between the active Cr-$e_{g}$ electrons via two different ligand $p$-orbitals, as schematically portrayed in Fig.\ref{SE_M}(c), benefits from distortion of CrTe$_6$ octahedra and Hund\textquoteright s energy gain at the ligand Te (or $X$) site to stabilize the FM order \cite{Zhang2019,Watson2019,Kang2019}. Moreover, correlation between $t_{2g}$-electrons also move up the ligand $p$ bands close to Fermi level, thus opening a band gap between Cr $d$ conduction band and ligand $p$ valence band confirming the charge transfer type Mott behaviour \cite{Suzuki2019b,Zhang2019,Kang2019,Watson2019,Kundu2020}. 

The charge gap of a Mott insulator can be manipulated either by carrier doping which fills the band or by controlling the bandwidth, resulting in an insulator-metal transition (IMT) often accompanied by a structural and magnetic transitions. So far, doping through gating a field effect transistor (FET) device \cite{Wang2018b,Jiang2018,Verzhbitskiy2020a}, or intercalation of organic ions to bulk single crystals \cite{Wang2019} have been used. With electrostatic gating to few layers of CrGeTe$_3$ \cite{Wang2018b} and CrI$_3$ \cite{Jiang2018}, the coercive field and saturation magnetization are found to be modulated, but $T_{\rm{C}}$ changes barely. However, when high carrier density ($\sim$ 10$^{-14}$ cm$^{-2}$) is doped to CrGeTe$_3$ either through intercalation \cite{Wang2019} or ionic liquid gating to FET devices \cite{Verzhbitskiy2020a}, $T_{\rm{C}}$ around $\sim$ 200 K is achieved with simultaneous stabilization of a metallic state. Although these results are promising, nevertheless, such filling controlled methods lead to undesirable effects such as strong charge in-homogeneity in atomic scale either due to disorder created by intercalant ions or chemical modification of the host material induced by ionic liquid \cite{Weber2019}.

On the contrary, application of pressure is an alternative route, which not only controls the bandwidth but also the spin exchange pathways via subtle modification of bond length and angle between atoms avoiding the complication of disorder. Pressure has been suitably used to switch the interlayer magnetism of few layers of CrI$_3$ \cite{Song2019,Li2019} including $T_{\rm{C}}$ variation of CrI$_3$ and CrGeTe$_3$ bulk single crystals \cite{Sun2018,Lin2018,Mondal}. However, these studies are limited to less than 2.0 GPa with $T_{\rm{C}}$ varying about $\sim$ 10\%. Interestingly, a recent high pressure work on CrSiTe$_3$ \cite{Cai2020} revealed a concomitant structural transition and IMT followed by a superconductivity around 7.0 GPa.

In this work, we study the electronic and magnetic properties of CrGeTe$_3$ single crystals by varying pressure up to 11.0 GPa using dc magnetic susceptibility and resistivity measurement. Although, the pressure dependent lattice dynamics of CrGeTe$_3$ have been performed revealing a 2D to 3D structural transition around $\ge$ 18 GPa \ \cite{Sun2018,Yu2019}, the magnetic and electrical properties at high pressure have remain unexplored. At ambient pressure, CrGeTe$_3$ is an insulator with a band gap of $\sim$ 0.7 eV \cite{Ji2013} and orders in a Heisenberg type ferromagnetism below $T_{\rm{C}} \sim$ 66 K as shown in Fig.\ref{SE_M}(d,e). At first,  $T_{\rm{C}}$ decreases monotonically to 54 K as pressure increases to 4.5 GPa. Remarkably, $T_{\rm{C}}$ jumps four-fold in between 4.5 $\leq P\leq$ 7.5 GPa and surpasses 250 K above 9.1 GPa. The pressure temperature phase diagram uncovers an IMT to a correlated Fermi metallic state above $\sim$ 7.0 GPa, characterized by large resistivity anisotropy, $\rho_{c}/\rho_{ab}\sim$10$^5$, suggesting 2D nature of charge transport. Our results suggest collapse of charge transfer energy gap between Te-$p$ valence band and Cr-$e_g$ conduction band dramatically boost the intra-layer FM superexchange interaction producing such a high $T_{\rm{C}}$. In addition, these results provide a rare example of bandwidth-controlled IMT in a vdW material, without entailing simultaneous structural transition and spin-crossover from magnetic to non-magnetic state.
\begin{figure}[th]
\includegraphics[width=0.47\textwidth]{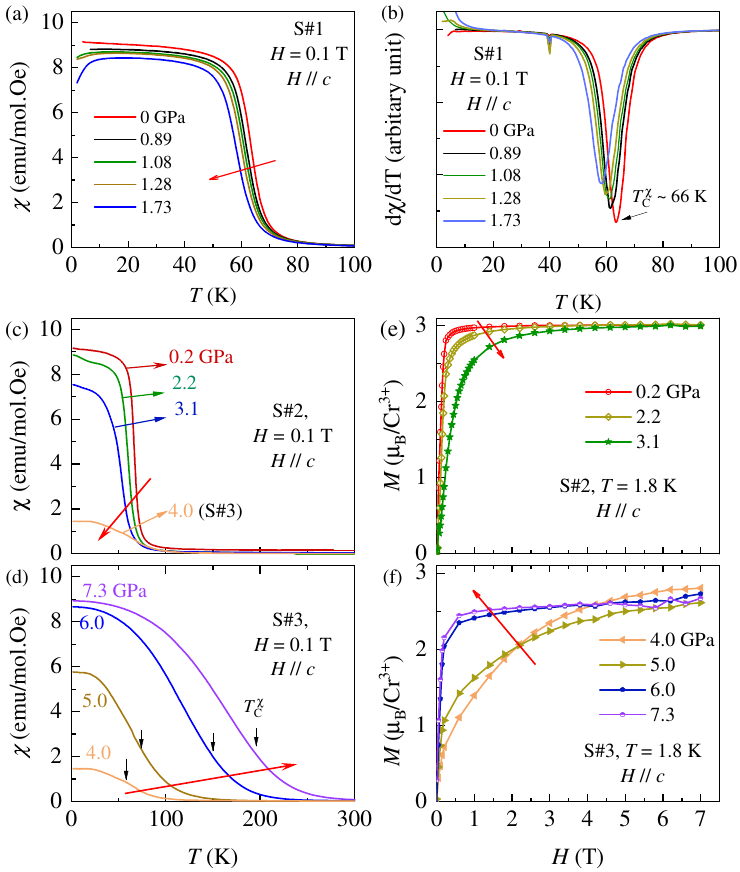}
\caption{(a) Temperature dependence of ZFC dc susceptibility, $\chi$ of CrGeTe$_3$ crystal S\#1 at $H$ = 0.1 T applied parallel to $c$-axis for pressure up to 1.73 GPa. (b) Temperature dependence of $d\chi/dT$ of S\#1. The temperature dependence of field-cooled (FC) susceptibility, $\chi$, for (c) crystal S\#2 and (d) crystal S\#3, respectively. The field dependence of magnetization, $M(H)$, for (e) S\#2 and (f) S\#3 at $T$ = 1.8 K under different pressure. Red arrows in figures represent the direction of increasing pressure.}\label{susc}
\end{figure}

In Fig.\ref{susc}(a), we illustrate the temperature dependence of dc susceptibility, $\chi$, of CrGeTe$_3$ single crystal (S\#1) with an applied magnetic field $H$ = 0.1 T parallel to $c$-axis at several pressure. At ambient pressure, $\chi$($T$) undergoes a typical paramagnet to FM transition and the estimated $T_{\rm{C}}^{\chi}$$\sim$ 66 K, from the minimum of $d\chi/dT$ curve is consistent with several earlier reports \cite{Liu2017a,Ji2013,Liu2018b}. With increasing pressure up to 1.73 GPa (Fig.\ref{susc}(a)), the transition shifts towards lower temperature and $T_{\rm{C}}^{\chi}$ reduces to $\sim$ 58 K at 1.73 GPa [see Fig.\ref{susc}(b)]. The weak downturn of $\chi$ below $\sim$ 7.0 K at 1.08 GPa is related to the formation of FM domains in the crystal as evidenced by the bifurcation between FC and ZFC curve [see supplementary Fig.S2]. To track further the evolution of $T_{\rm{C}}$ above 2.0 GPa, we measured $\chi$ of another two pieces of CrGeTe$_3$ single crystals extracted from the same batch using an opposed-anvil type pressure cell \cite{Hiraoka2020}. Fig.\ref{susc}(c) and (d) present the temperature dependence of $\chi$ for crystals S\#2 and S\#3, respectively. Although the magnitude of $\chi$ decreases drastically as pressure is raised to 4.0 GPa [Fig.\ref{susc}(c)], $T_{\rm{C}}^{\chi}$ decreases modestly to $\sim$ 54 K. Surprisingly, as pressure is increased further the magnitude of $\chi$ surges again and transition shifts towards higher temperature [Fig.\ref{susc}(d)]. For $P\ge$ 5.0 GPa, $\chi(T)$ curves show a broad transition to a FM state with $T_{\rm{C}}^{\chi} \sim$ 73 K. At 7.3 GPa, $T_{\rm{C}}^{\chi}$ reaches as high as $\sim$ 196 K, almost three times higher than the $T_{\rm{C}}^{\chi}$ at ambient pressure.

In Fig.\ref{susc}(e) and (f), we present the $M$($H$) curve along the easy magnetization $c$-axis for crystals S\#2 and S\#3, respectively. At 0.2 GPa, $M$($H$) curve saturates sharply at $H_{\rm{S}}\cong$ 0.22 T with a saturation moment $M_{\rm{S}}\sim 3.0 \mu_B$/Cr$^{3+}$. With variation of pressure $M$($H$) curve becomes rounder and $H_{\rm{S}}$ surpasses 3.0 T at 3.1 GPa. At 4.0 GPa, $M$($H$) curve look similar to a Brillouin like function and does not saturate even up to 7.0 T. Even though, $M$($H$) curve resembles typically observed in a paramagnet, the Curie-Weiss temperature, $\theta_{\rm{CW}}$, at this pressure is comparable with that of ambient condition, hinting the presence of strong FM interaction (see supplementary Fig.S3). For $P\geq$ 5.0 GPa, $H_{\rm{S}}$ start to decrease again. At $P\cong$ 6.0 GPa, $M$($H$) curve saturate sharply above $H_{\rm{S}}\simeq$ 0.18 T with $M_{\rm{S}}\sim$ 2.76 $\mu_B$/Cr$^{3+}$, a typical sign of FM order. From the evolution of $\chi$($T$) and $M$($H$) curves under pressure, it is clear that in the intermediate pressure range 3.1 GPa $\le P \le$ 5.0 GPa, the net FM exchange interaction weakens, producing a diminished $\chi$($T$) magnitude and $T_{\rm{C}}^{\chi}$.

\begin{figure}[th]
\includegraphics[width=0.47\textwidth]{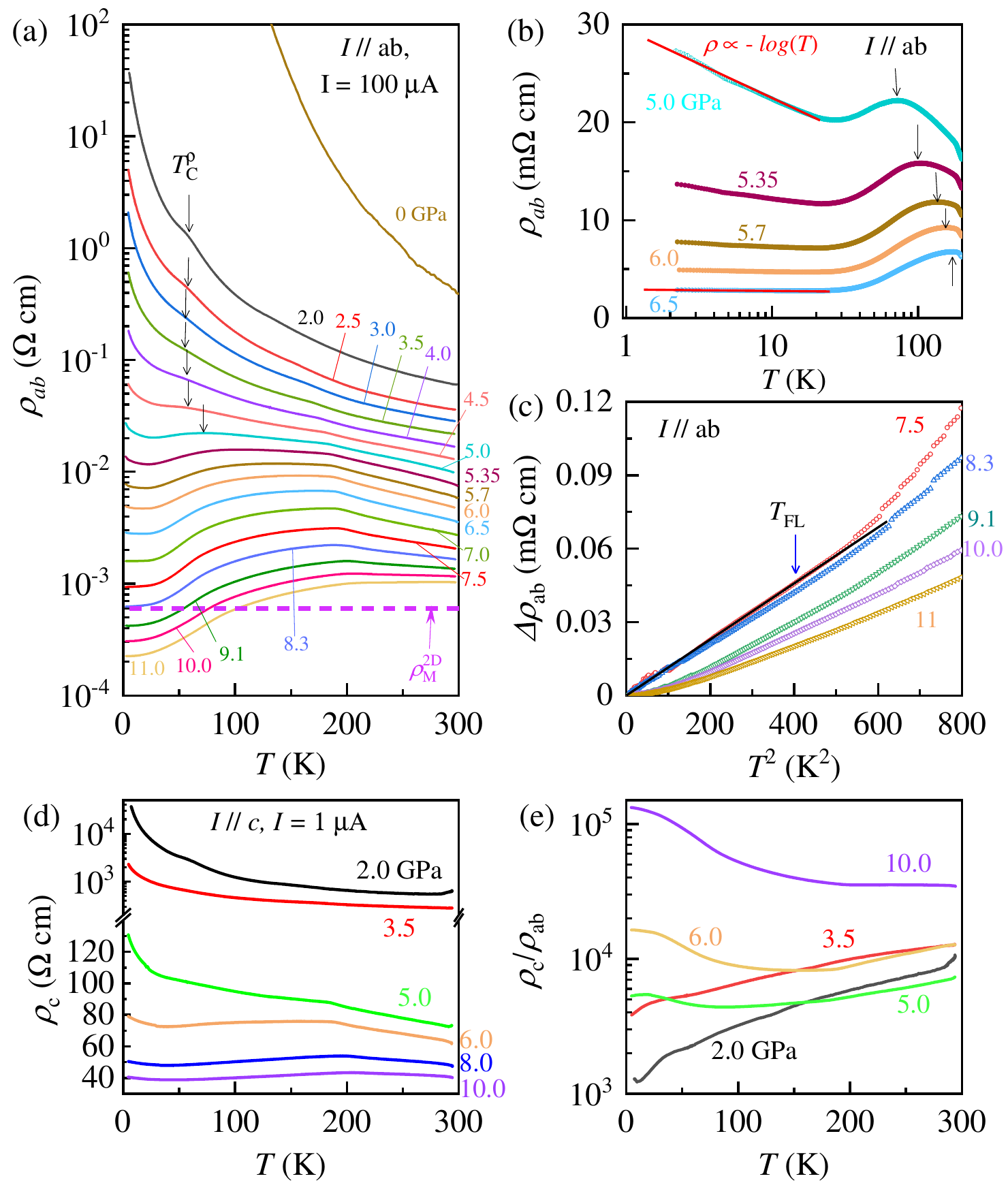}
\caption{(a) Temperature dependence of in-plane resistivity, $\rho_{ab}$, of a CrGeTe$_3$ single crystal at several pressure ranging from 0 to 11.0 GPa, disclosing an insulator to metal transition. The horizontal magenta dotted line represents an estimate of the Mott-Ioffe-Regel limit of resistivity, $\rho_{\rm{M}}^{\rm{2D}}$, as described in the text. (b) Temperature dependence of $\rho_{ab}$ in logarithmic temperature scale. Red solid lines are -$\log(T)$ fit to $\rho_{ab}$. Black downward arrows trace $T_c^{\rho}$ with pressure. (c) $\Delta\rho_{ab}$ = ($\rho_{ab}-\rho_{0}$) vs $T^2$ for $P \ge$ 7.5 GPa, where $\rho_{0}$ is the residual resistivity. The black line is a linear fit to $\Delta\rho$ at 7.5 GPa confirming a Fermi liquid state below $T_{\rm{FL}}$. (d) The out-of-plane resistivity, $\rho_{c}$, at some selected pressure. (e) Temperature dependence of resistivity anisotropy, $\rho_{c}$/$\rho_{ab}$, at different pressure.}\label{res}
\end{figure}

Next, we show the temperature dependence of in-plane resistivity, $\rho_{ab}$, of CrGeTe$_3$ under pressure ranging from 0 to 11 GPa, in Fig.\ref{res}(a), revealing an insulator to metal transition (IMT). The overall $\rho_{ab}$ plummet several orders of magnitude as 11.0 GPa pressure is applied; at low temperature $\rho_{ab}$ falls more than nine orders of magnitude, while at $T$ = 300 K more than three orders. At 2.0 GPa, $\rho_{ab}$($T$) discloses a shoulder-like anomaly at $T_{\rm{C}}^{\rho}\sim$ 59 K, very close to the FM transition observed in $\chi$($T$), allowing to trace the evolution of magnetic order with pressure from $\rho_{ab}$ data. $T_{\rm{C}}^{\rho}$ falls to $\sim$ 54 K at 4.5 GPa. However, at 5.0 GPa, $\rho_{ab}$ exhibits a broad maximum at $\sim$ 73 K [Fig.\ref{res}(b)], an indication of $T_{\rm{C}}^{\rho}$ moving to high temperature in agreement with the $T_{\rm{C}}^{\chi}$ of $\chi$($T$). Additionally, $\rho_{ab}$ reveals a weak upturn below $\sim$ 25 K, where it follows a -$\log$($T$) dependence with decreasing temperature suggesting a 2D weak localization behavior. On compressing further to 7.0 GPa, $T_{\rm{C}}^{\rho}$ surges rapidly to 193 K consistent with $\chi$($T$) data. At the same time, -$\log$($T$) upturn in $\rho_{ab}$ is gradually suppressed uncovering a metallic state. At high pressure, precise determination of $T_{\rm{C}}^{\rho}$ becomes difficult due to the broad anomaly, yet signature of $T_{\rm{C}}^{\rho}\sim$ 250 K can be traced up to 9.1 GPa. 

At 7.0 GPa, $\rho_{ab}$ follows a power-law, $\rho_{ab} = \rho_{0} +AT^n$ with an exponent $n\sim$ 1.9 below $T_{\rm{FL}}$$\sim$ 14 K, a hallmark of Fermi-liquid (FL) state. It is worthy to note that at this pressure, $\rho_{ab}$ is higher than the Mott-Ioffe-Regel limit of resistivity defined as $\rho_{\rm{M}}^{\rm{2D}}\cong$ 0.055($c/a_0$)$\sim$ 0.6 m$\Omega$ cm, where $c$ is the average separation between CrGeTe$_3$ layers and $a_0$ = 0.529 $\AA$ is the Bohr radius \cite{Gunnarsson2003}. Aside, the residual resistivity ratio (RRR), $\rho_{300 \rm{K}}/\rho_{2 \rm{K}}\simeq$ 2 clearly indicates a bad metal behavior. The RRR and temperature window, where $\rho_{ab} < \rho_{\rm{M}}^{\rm{2D}}$, increases with rising pressure. Simultaneously, $T_{\rm{FL}}$ grows to 56 K at 11.0 GPa [Fig.\ref{res}(c)] and $A$-coefficient of $T^2$-term drops by more than two orders of magnitude, a clear indication of widening of the FL region at higher pressure with decreasing correlation strength. 

To obtain more insight about this metallic state, we measured inter-layer resistivity, $\rho_{c}$, of another piece of single crystal from the same batch as shown in Fig.\ref{res}(d). Unlike $\rho_{ab}$, $\rho_{c}$ continue to be semiconducting down to the lowest temperature at 6.0 GPa. Intriguingly, $\rho_{c}$ is nearly temperature independent deep inside the metallic state even at 10.0 GPa, conveying an incoherent inter-layer charge transport and confinement of charge carriers in the $ab$-plane. This  becomes increasingly clear from the temperature dependence of resistivity anisotropy, $\rho_{c}/\rho_{ab}$ as in shown Fig.\ref{res}(e). For $P<$ 5.0 GPa, $\rho_{c}/\rho_{ab}$ $\simeq$ 10$^3$-10$^4$, whereas $\rho_{c}/\rho_{ab}$ reaches as high as $\sim$ 10$^5$ at 10.0 GPa and 2 K. Such a large value of $\rho_{c}/\rho_{ab}$ is comparable with that of several strongly correlated materials like high-$T_c$ cuprates \cite{Ono2003}, manganites \cite{Kuwahara1999} and organic compounds \cite{Shimizu2018}. Thus, the pressure induced metallic state in CrGeTe$_3$ can be regarded as an ideal 2D correlated Fermi metal. 

\begin{figure}[th]
\includegraphics[width=0.47\textwidth]{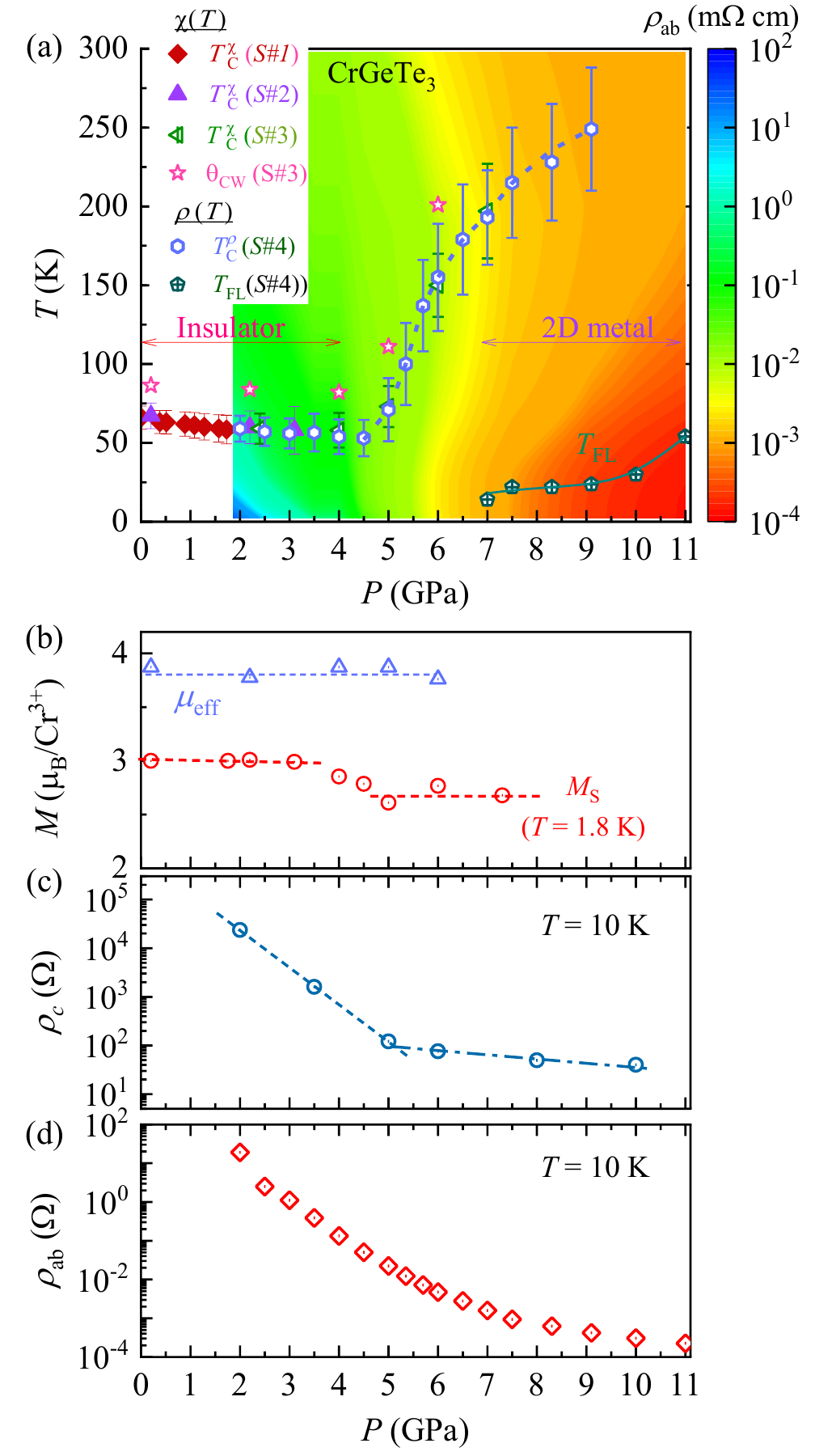}
\caption{(a) Pressure temperature phase diagram of CrGeTe$_3$. Color scale represents the magnitude of $\rho_{ab}$. $\theta_{\rm{CW}}$ and $T_{\rm{C}}^{\chi}$ respectively are the Curie-Weiss temperature and Curie temperature, estimated from magnetic susceptibility. $T_{\rm{C}}^{\rho}$ and $T_{\rm{FL}}$ are Curie temperature and Fermi liquid temperature determined from $\rho_{ab}$. (b) Pressure dependence of effective moment, $\mu_{\rm{eff}}$, and saturation magnetic moment $M_{\rm{S}}$ (at $T$ = 1.8 K). (c) and (d) pressure dependence of inter-layer resistivity, $\rho_{c}$, and in-plane resistivity $\rho_{ab}$ at $T$ = 10 K, respectively.}\label{pd}
\end{figure}

The $P$-$T$ phase diagram [Fig.\ref{pd}(a)], constructed from susceptibility and resistivity data, unveils a fascinating evolution of the electronic and magnetic properties of CrGeTe$_3$ when pressure is continuously varied. With adjustment of pressure, CrGeTe$_3$ alter from a insulator with $T_{\rm{C}} \sim$ 66 K at ambient pressure to a FMM with $T_{\rm{C}}$ surpassing 250 K. The moderate decrease of $T_{\rm{C}}$ to $\sim$ 54 K as pressure is raised to 4.5 GPa and subsequent dramatic four-fold rise of $T_{\rm{C}}$ in a rather narrow pressure range 4.5 $\leq P\leq$ 6.5 GPa emphasizes the competition between FMI and FMM phase. The phase diagram also unearth a correlated 2D Fermi metal for $P \ge$7.0 GPa, significantly well below the 2D to 3D structural phase transition pressure $\sim$ 18 GPa \cite{Yu2019}, revealing a contrasting feature of CrGeTe$_3$ with the CrSiTe$_3$ \cite{Cai2020} and isostructural $M$P$X_3$ ($M$ = V, Fe, Mn, Ni and $X$ = S, Se) compounds. 

When pressurized, $M$P$X_3$ undergo an IMT accompanied by an iso-symmetric structural transition, forming a $M$-$M$ dimer due to direct overlap between neighboring $t_{2g}$ orbitals of $M$ ions \cite{Wang2016c,Haines2018,Wang2018,Coak2019,Kim2019b}. Therefore, a spin-crossover from high-spin to low-spin state occurs together with a large volume collapse. On the other hand, the non-existence of such lattice change close to IMT of CrGeTe$_3$ \cite{Sun2018,Yu2019}, assert that direct interaction between $t_{2g}$- electrons of neighboring Cr atoms is still weak. In fact, this is compatible with the nearly pressure independence of $M_{\rm{S}}$ affirming the survival of localized high spin ($S$ = 3/2) state of $t_{2g}$-manifolds even in high pressure state [Fig.\ref{pd}(b)]. Rather, these observations hint the active Cr-$e_g$ orbital plays major role in stabilizing the high pressure metallic state in CrGeTe$_3$, thus averting a structural transition as recently proposed for NiPS$_3$ (Ref.\cite{Kim2019b}). Moreover, the remarkable absence of concomitant structural distortion across the IMT signify that the pressure induced modification of magnetic and electronic ground states of CrGeTe$_3$ are purely electronic in origin. 

Having discussed the phase diagram, now we qualitatively explain the origin of such intriguing pressure dependent $T_{\rm{C}}$ of CrGeTe$_3$. For a ferromagnet in 2D limit, the MAE, $K$, along with the spin-exchange interaction, $J$, determine $T_{\rm{C}}\sim J/ln(3\pi J/4K)$ \cite{Review1988}. For CrGeTe$_3$, $K$ ($\simeq$ -0.05 meV), usually set by SOC and influence the direction of magnetic easy axis \cite{Gong2017, Kim2019}, is much smaller in energy scale compared to $J$ ($\simeq$ 3.0 meV) \cite{Gong2017,Fumega2020a,Menichetti2019}. Thus, the strength of exchange interaction plays a major role in fixing the $T_{\rm{C}}$. As explained earlier in Fig.\ref{SE_M}(c), the FM order in CrGeTe$_3$ is mainly driven by indirect superexchange interaction, $J_{\rm{SE}}$, between Cr-$e_{g}$ orbitals via two orthogonal Te-$p$ orbitals. With such orbital arrangement, the superexchange integral can be written as \cite{Streltsov2008,Khomskii2016}
\begin{equation}
	J_{\rm{SE}}\sim -\frac{t_{pd}^{2}t_{p'd'}^{2}J_{\rm{H}}^{\rm{Te}}}{\Delta_{\rm{CT}}^2(2\Delta_{\rm{CT}}+U_{p})^2},
	\label{JSE}
\end{equation}
where $t_{pd}(t_{p'd'})$ is the virtual hopping between $e_{g}$($e_g'$) and $p$($p'$) orbital, $J_{\rm{H}}^{\rm{Te}}$ is the Hund\textquoteright s coupling on Te ion, $U_{p}$ is the Coulomb interaction on Te site and $\Delta_{\rm{CT}}$ is the charge transfer energy gap between Te-$p$ valence band top and bottom of the Cr-$e_g$ conduction band. Also, $J_{\rm{SE}}$ depends on the geometrical Cr-Te-Cr bond angle $\theta$. Pressure dependent x-ray diffraction studies reveal that $\theta$ deviates marginally from 90$^{\circ}$ under pressure even up to 10.0 GPa \cite{Sun2018,Fumega2020a,Yu2019} and with such small deviation $J_{\rm{SE}}$ exchange path remains robust \cite{Dong2020a}. Assuming $J_{\rm{H}}^{\rm{Te}}$ to be unchanged with pressure, from Eq.(\ref{JSE}) it is evident that decrease of $\Delta_{\rm{CT}}$ will have much stronger impact on $J_{\rm{SE}}$ compared to others, since, $\Delta_{\rm{CT}}$ can be several times larger than $t_{pd}$ and $U_{p}$. It is credible to expect $\Delta_{\rm{CT}}$ tends to zero on approaching IMT with varying pressure. Indeed, a recent first principle calculation predicts the collapse of $p$-$d$ energy gap around 7.0 GPa as a result of shortened Cr-Te bond \cite{Fumega2020a}. Therefore, the sharp rise of $T_{\rm{C}}$ for $P\ge$ 4.5 GPa can be attributed to the dramatic upsurge of intra-layer $J_{\rm{SE}}$ resulting from the depreciation of $\Delta_{\rm{CT}}$ with pressure.

The initial decrease of $T_{\rm{C}}$ and $\chi$ up to 4.5 GPa could be attributed to the following reasons. At ambient condition, apart from $J_{\rm{SE}}$, there exists finite AFM exchange interactions between the next nearest neighbor Cr atoms in the layer as well as between the Cr atoms in adjacent layers \cite{Gong2017,Menichetti2019,Fumega2020a,Zhang2019}. With growing pressure these AFM interactions will compete with $J_{\rm{SE}}$, as they are expected to strengthen due to shrinking distance between Cr atoms and vdW gaps between adjacent layers. Another is the sign change of $K$ from negative to positive with pressure, that will trigger an alteration of magnetic easy axis from $c$-axis to $ab$-plane as reported in pressure dependent magnetoresistance study \cite{Lin2018}. However, when pressure exceeds 4.5 GPa, $J_{\rm{SE}}$ overcomes this competition due to depreciating charge gap with pressure. The distinct pressure dependence of $\rho_c$ and $\rho_{ab}$, respectively displayed in Fig.\ref{pd}(c) and (d), back up the dominance of intra-layer $J_{\rm{SE}}$ at high pressure region. Up to 5.0 GPa, $\rho_c$ drops more than two orders of magnitude, whereas it decreases barely in between 6.0 to 10.0 GPa. By contrast, $\rho_{ab}$ continue to plunge up to 11.0 GPa. As well, the large $\rho_{c}/\rho_{ab}$ $\sim$ 10$^5$ in high pressure metallic state points that carriers can hop more easily inside the layer rather than across the layers, making the intra-layer $J_{\rm{SE}}$ significantly stronger than inter-layer interactions. 

In summary, we demonstrate that pressure can be used as a suitable parameter to control both magnetic and electrical properties of CrGeTe$_3$ by tuning the charge transfer energy gap between Te $p$ valence band and Cr $e_g$ conduction band. Moreover, lack of concurrent structural transition and spin-crossover across the IMT makes CrGeTe$_3$ an unique vdW material and provides a novel example of bandwidth-controlled IMT. Our results also indicate that the electronic properties of CrGeTe$_3$ can be much more responsively switched compared to other vdW materials, by external tuning parameter such as doping, thin film strain or electrostatic gating. 

We acknowledge fruitful discussions with J.-G. Cheng, K. Matsubayashi, M. K. Ray and K. Murata. This work was financially supported by the JSPS KAKENHI Grant numbers JP19H00648 and 19H01836. The work at SNU was supported by grant no. 2019R1A2C2090648, 2019M3E4A1080227 and 2016K1A4A3914691 through the National Research Foundation of Korea funded by the Korean government.


\begin{thebibliography}{99}
\bibitem{Burch2018} K. S. Burch, D. Mandrus and J.-G. Park, Nature, \textbf{563}, 47-52 (2018).

\bibitem{Huang2020} B. Huang, M. A. McGuire, A. F. May, D. Xiao, P. Jarillo-Herrero and X. Xu, Nat. Mater., \textbf{19}, 1276 (2020).

\bibitem{Gong2019} C. Gong and X. Zhang, Science \textbf{363}, eaav4450 (2019).

\bibitem{Zhong2017} D. Zhong, K. L. Seyler, X. Linpeng, R. Cheng, N. Sivadas, B. Huang, E. Schmidgall, T. Taniguchi, K. Watanabe, M. A. McGuire \textit{et al.}, Sci. Adv. \textbf{3}, e1603113 (2017).

\bibitem{Yao2019} X. Yao, B. Gao, M.-G. Han, D. Jain, J. Moon, J. W. Kim, Y. Zhu, S.-W. Cheong and S. Oh, Nano Lett, \textbf{19}, 4567 (2019).

\bibitem{Seyler2018} K. L. Seyler, D. Zhong, B. Huang, X. Linpeng, N. P. Wilson, T. Taniguchi, K. Watanabe, W. Yao, D. Xiao, M. A. McGuire, Kai-Mei C. Fu \textit{et al.}, Nano Lett., \textbf{18}, 3823 (2018).

\bibitem{Xu2020} C. Xu, J. Feng, M. Kawamura, Y. Yamaji, Y. Nahas, S. Prokhorenko, Y. Qi, H. Xiang, and L. Bellaiche, Phys. Rev. Lett. \textbf{124}, 087205 (2020).

\bibitem{Xu2019} C. Xu, J. Feng, H. Xiang and L. Bellaiche, npj Comput. Mater. \textbf{4}, 57 (2018).

\bibitem{Gong2017} C. Gong, L. Li, Z. Li, H. Ji, A. Stern, Y. Xia, T. Cao, W. Bao,
C. Wang, Y. Wang \textit{et al.}, Nature, \textbf{546}, 265 (2017).

\bibitem{Carteaux1995} V. Carteaux, D. Brunet, G. Ouvrard and G. Andre \textit{et al.}, J. Phys. Condens. Matter, \textbf{7}, 69 (1995).

\bibitem{McGuire2015} M. A. McGuire, H. Dixit, V. R. Cooper and B. C. Sales, Chem. Mater. \textbf{27}, 612 (2015).

\bibitem{Tsubokawa} I. Tsubokawa, J. Phys. Soc. Japan \textbf{15}, 1664 (1960).

\bibitem{Huang2017} B. Huang, G. Clark, E. Navarro-Moratalla, D. R. Klein, R. Cheng, K. L. Seyler, D. Zhong, E. Schmidgall, M. A. McGuire \textit{et al.}, Nature,\textbf{ 546}, 270 (2017).

\bibitem{Zhang2019a} Z. Zhang, J. Shang, C. Jiang, A. Rasmita, W. Gao, and T. Yu, Nano. Lett. \textbf{19}, 3138 (2019).

\bibitem{Kim2019c} H. H. Kim, B. Yang, S. Li, S. Jiang, C. Jin, Z. Tao, G. Nichols, F. Sfigakis, S. Zhong, C. Li, S. Tian \textit{et al.}, Proc. Natl. Acad. Sci. \textbf{116}, 11131 (2019).

\bibitem{Freitas2005} D. C. Freitas, R. Weht, A. Sulpice, G. Remenyi, P. Strobel, F. Gay, J. Marcus, and M. N.-Regueiro, J. Phys.: Condens. Matter \textbf{27}, 176002 (2015).

\bibitem{Zhang2021} X. Zhang, Q. Lu, W. Liu, W. Niu, J. Sun, J. Cook, M. Vaninger, P. F. Miceli, D. J. Singh, S.-W. Lian \textit{et al.}, Nat. Commun. \textbf{12}, 2492 (2021).

\bibitem{Deiseroth2006} H. J. Deiseroth, K. Aleksandrov, C. Reiner, L. Kienle, and R. K. Kremer, Eur. J. Inorg. Chem. \textbf{2006}, 1561 (2006).

\bibitem{Chen2013} B. Chen, J. Yang, H. D. Wang, M. Imai, H. Ohta, C. Michioka, K. Yoshimura, and M. Fang, J. Phys. Soc. Japan \textbf{82}, 124711 (2013).

\bibitem{Deng2018} Y. Deng, Y. Yu, Y. Song, J. Zhang, N. Z. Wang, Z. Sun, Y. Yi, Y. Z. Wu, S. Wu, J. Zhu \textit{et al.}, Nature \textbf{563}, 94 (2018).

\bibitem{Fei2018}  Z. Fei, B. Huang, P. Malinowski, W. Wang, T. Song, J. Sanchez, W. Yao, Di Xiao, X. Zhu, A. F. May \textit{et al.}, Nat. Mater. \textbf{17}, 778 (2018).

\bibitem{Kang2019} S. Kang, S. Kang, and J. Yu, J. Electron. Mater. \textbf{48}, 1441 (2019).

\bibitem{Suzuki2019b} M. Suzuki, B. Gao, K. Koshiishi, S. Nakata, K. Hagiwara, C. Lin, Y. X. Wan, H. Kumigashira, K. Ono, S. Kang \textit{et al.}, Phys. Rev. B, \textbf{99}, 161401(R) (2019).

\bibitem{Zhang2019} J. Zhang, X. Cai, W. Xia, A. Liang, J. Huang, C. Wang, L. Yang, H. Yuan, Y. Chen, S. Zhang \textit{et al.}, Phys. Rev. Lett. \textbf{123}, 047203 (2019).

\bibitem{Kim2019} D.-H. Kim, K. Kim, K.-T. Ko, J. Seo, J. S. Kim, T.-H. Jang, Y. Kim, J.-Y. Kim, S.-W. Cheong, and J.-H. Park, Phys. Rev. Lett. \textbf{122}, 207201 (2019).

\bibitem{Watson2019} M. D. Watson, I. Markovi\'{c}, F. Mazzola, A. Rajan, E. A. Morales, D. M. Burn, T. Hesjedal, G. van der Laan, S. Mukherjee, T. K. Kim \textit{et al.}, Phys. Rev. B \textbf{101}, 205125 (2020).

\bibitem{Kundu2020} A. K. Kundu, Y. Liu, C. Petrovic, and T. Valla, Sci. Rep. \textbf{10}, 15602 (2020).

\bibitem{Wang2018b} Z. Wang, T. Zhang, M. Ding, B. Dong, Y. Li, M. Chen, X. Li, J. Huang, H. Wang, X. Zhao \textit{et al.}, Nat. Nanotechnol. \textbf{13}, 554 (2018).

\bibitem{Jiang2018} S. Jiang, L. Li, Z. Wang, K. F. Mak, and J. Shan, Nat. Nanotechnol. \textbf{13}, 549-553 (2018).

\bibitem{Verzhbitskiy2020a} I. A. Verzhbitskiy, H. Kurebayashi, H. Cheng, J. Zhou, S. Khan, Y. P. Feng, and G. Eda, Nat. Electron., \textbf{3}, 460 (2020).

\bibitem{Wang2019} N. Wang, H. Tang, M. Shi, H. Zhang, W. Zhuo, D. Liu, F. Meng, L. Ma, J. Ying, L. Zou \textit{et al.}, J. Am. Chem. Soc., \textbf{141}, 17166 (2019).

\bibitem{Weber2019} D. Weber, A. H. Trout, D. W. McComb, and J. E. Goldberge, Nano Lett. \textbf{19}, 5031 (2019).

\bibitem{Song2019} T. Song, Z. Fei, M. Yankowitz, Z. Lin, Q. Jiang, K. Hwangbo, Q. Zhang, B. Sun, T. Taniguchi, K. Watanabe \textit{et al.}, Nat. Mater., \textbf{18}, 1298 (2019).

\bibitem{Li2019} T. Li, S. Jiang, N. Sivadas, Z. Wang, Y. Xu, D. Weber, J. E. Goldberger, K. Watanabe, T. Taniguchi, C. J. Fennie \textit{et al.}, Nat. Mater. \textbf{18}, 1303 (2019).

\bibitem{Sun2018} Y. Sun, R. C. Xiao, G. T. Lin, R. R. Zhang, L. S. Ling, Z. W. Ma, X. Luo, W. J. Lu, Y. P. Sun, and Z. G. Sheng,  Appl. Phys. Lett., \textbf{112}, 072409 (2018).

\bibitem{Lin2018} Z. Lin, M. Lohmann, Z. A. Ali, C. Tang, J. Li, W. Xing, J. Zhong, S. Jia, W. Han, S. Coh \textit{et al.}, Phys. Rev. Mater. \textbf{2}, 051004(R) (2018).

\bibitem{Mondal} S. Mondal, M. Kannan, M. Das, L. Govindaraj, R. Singha, B. Satpati,
S. Arumugam, and P. Mandal, Phys. Rev. B \textbf{99}, 180407(R) (2019).

\bibitem{Cai2020} W. Cai, H. Sun, W. Xia, C. Wu, Y. Liu, H. Liu, Y. Gong, D.-X. Yao, Y. Guo, and M. Wang, Phys. Rev. B \textbf{102}, 144525 (2020).

\bibitem{Yu2019} Z. Yu, W. Xia, K. Xu, M. Xu, H. Wang, X. Wang, N. Yu, Z. Zou, J. Zhao, L. Wang, X. Miao, and Y. Guo, J. Phys. Chem. C, \textbf{123}, 13885 (2019).

\bibitem{Ji2013} H. Ji, R. A. Stokes, L. D. Alegria, E. C. Blomberg, M. A. Tanatar, A. Reijnders, L. M. Schoop, T. Liang, R. Prozorov, K. S. Burch \textit{et al.}, J. Appl. Phys., \textbf{114}, (2013).

\bibitem{Liu2017a} Y. Liu and C. Petrovic, Phys. Rev. B, \textbf{96}, 054406 (2017).

\bibitem{Liu2018b} W. Liu, Y. Dai, Y.-E. Yang, J. Fan, L. Pi, L. Zhang, and Y. Zhang, Phys. Rev. B, \textbf{98}, 214420 (2018).

\bibitem{Hiraoka2020} N. Hiraoka, K. Whiteaker, M. Blankenhorn, Y. Hayashi, R. Oka, H. Takagi, and K. Kitagawa, J. Phys. Soc. Jpn. \textbf{90}, 074001 (2021)

\bibitem{Gunnarsson2003} O. Gunnarsson, M. Calandra, and J. E. Han, Rev. Mod. Phys. \textbf{75}, 1085 (2003).

\bibitem{Ono2003} S. Ono, and Y. Ando, Phys. C Supercond., \textbf{388-389}, 321 (2003).

\bibitem{Kuwahara1999} H. Kuwahara, T. Okuda, Y. Tomioka, A. Asamitsu, and Y. Tokura, Phys. Rev. Lett., \textbf{82}, 4316-4319 (1999).

\bibitem{Shimizu2018} Y. Shimizu, and R. Kato, Phys. Rev. B, \textbf{97}, 125107 (2018).

\bibitem{Wang2016c} Y. Wang, Z. Zhou, T. Wen, Y. Zhou, N. Li, F. Han, Y. Xiao, P. Chow, J. Sun, M. Pravica \textit{et al.}, J. Am. Chem. Soc., \textbf{138}, 15751 (2016).

\bibitem{Haines2018} C. R. S. Haines, M. J. Coak, A. R. Wildes, G. I. Lampronti, C. Liu, P. Nahai-Williamson, H. Hamidov, D. Daisenberger, and S. S. Saxena, Phys. Rev. Lett., \textbf{121}, 266801 (2018)

\bibitem{Wang2018} Y. Wang, J. Ying, Z. Zhou, J. Sun, T. Wen, Y. Zhou, N. Li, Q. Zhang, F. Han, Y. Xiao \textit{et al.}, Nat. Commun., \textbf{9}, 1914 (2018).

\bibitem{Coak2019} M. J. Coak, S. Son, D. Daisenberger, H. Hamidov, C. R. S. Haines, P. L. Alireza, A. R. Wildes, C. Liu, S. S. Saxena, and Je-Geun Park, npj Quantum Mater., \textbf{4}, 38 (2019).

\bibitem{Kim2019b} H. S. Kim, K. Haule, and D. Vanderbilt, Phys. Rev. Lett., \textbf{123}, 236401 (2019).

\bibitem{Review1988} Myron Bander and D. L. Mills, Phys. Rev. B, \textbf{38}, 12015(R) (1988). 

\bibitem{Fumega2020a} A. O. Fumega, S. Blanco-Canosa, H. Babu-Vasili, P. Gargiani, H. Li,  J.-S. Zhou, F. Rivadulla, and V. Pardo, J. Mater. Chem. C, \textbf{8}, 13582-13589 (2020).

\bibitem{Menichetti2019} G. Menichetti, M. Calandra, and M. Polini, 2D Mater., \textbf{6}, 045042 (2019).

\bibitem{Streltsov2008} S. V. Streltsov and D. I. Khomskii, Phys. Rev. B, \textbf{77}, 064405 (2008).

\bibitem{Khomskii2016} D. I. Khomskii, K. I. Kugel, A. O. Sboychakov, and S. V. Streltsov, J. Exp. Theor. Phys., \textbf{122}, 484 (2016).

\bibitem{Dong2020a} X. J. Dong, J. Y. You, Z. Zhang, B. Gu, and G. Su, Phys. Rev. B, \textbf{102}, 144443 (2020).

\end{thebibliography}
\end{document}